\documentclass[aps,prb,superscriptaddress,showpacs]{revtex4}
\usepackage{graphics}
\usepackage{epsfig}
\usepackage{graphicx}
\usepackage{dcolumn}

\begin{document}

\title{Strong spin relaxation length dependence on electric field gradients}

\author{D. Csontos}
\email{d.csontos@massey.ac.nz}
\affiliation{Institute of Fundamental Sciences, Massey University,
Private Bag 11222, Palmerston North, New Zealand}
\affiliation{Nanoscale and Quantum Phenomena Institute,
and Department of Physics and Astronomy, Ohio University, Athens, Ohio 45701-2979, USA}
\author{S.E. Ulloa}
\affiliation{Nanoscale and Quantum Phenomena Institute,
and Department of Physics and Astronomy, Ohio University, Athens, Ohio 45701-2979, USA}

\begin{abstract}
We discuss the influence of electrical effects on spin transport,
and in particular the propagation and relaxation of spin polarized
electrons in the presence of inhomogeneous electric fields. We show
that the spin relaxation length strongly depends on electric field
gradients, and that significant suppression of electron spin
polarization can occur as a result thereof. A discussion in terms of
a drift-diffusion picture, and self-consistent numerical
calculations based on a Boltzmann-Poisson approach shows that the
spin relaxation length in fact can be of the order of the charge
screening length.
\end{abstract}
\maketitle                   

\section{Introduction}
{\em Electrical} spin injection, transport and detection in diluted magnetic
semiconductor (DMS)-nonmagnetic semiconductor (NMS) based systems 
\cite{fiederlingohno}-\cite{kohdaCONDMAT2005} are three
crucial prerequisites for the successful realization of
semiconductor spintronics. While the spin degree of freedom naturally is important,
the charge degree of freedom plays a significant role in
semiconductor spin transport as well, through electric field induced
effects. Several experiments\cite{schmidtPRL2004}-\cite{crookerPRL2005}
have recently
shown that the spin injection efficiency is strongly dependent on
both applied and intrinsic electric fields, caused by, e.g.,
inhomogeneous doping. This was theoretically emphasized by Yu and
Flatte(YF)\cite{yu} within a drift-diffusion model from
which it was shown that the magnitude and sign of an electric field
strongly influences the spin relaxation length according to
\begin{equation}
L_{D(U)}=\left\{ -(+)\frac{\left\vert eE\right\vert
}{2k_{B}T}+\sqrt{\left(
\frac{eE}{2k_{B}T}\right)^{2}+\frac{1}{L_{s}^{2}}}\right\} ^{-1}~,
\label{YF}
\end{equation} where $E$ is a homogeneous electric field,
and $L_{s}$ is an intrinsic (electric field-independent) spin
diffusion length. According to eq. (\ref{YF}), the electric field
enhances(suppresses) the spin relaxation length $L_{D}(L_{U})$ in
the direction anti-parallel(parallel) to the direction of the
electric field. However, the underlying assumption leading up to eq.
(\ref{YF}) was {\em local charge neutrality}, and thus homogeneous
electric fields. The question is how spin polarized electrons
propagate in, e.g., inhomogeneously doped semiconductors where
electric fields are inhomogeneous?

In this paper, we use a similar approach as YF as well as
self-consistent numerical calculations to study the effects of
inhomogeneous electric fields. We find that, in an analogous
drift-diffusion picture, a "quasi-local" spin relaxation length in
the presence of inhomogeneous electric fields can be defined
according to
\begin{equation}
L_{D(U)}^{\prime}=\left [-(+)\frac{|eE|}{2k_{B}T} + \sqrt{\left (
\frac{eE}{2k_{B}T}\right )^{2} + \frac{1}{L_{s}^{2}}- \frac{e\nabla
E}{k_{B}T} } \right ]^{-1}~,
\label{LDU}
\end{equation}
from which a strong electric field gradient dependence becomes
evident. However, in order to understand the full influence of
inhomogeneous electric fields one needs to have access to the
self-consistent charge density and electric field profile. We have
performed such self-consistent numerical calculations using a
Boltzmann-Poisson equation approach for a DMS-NMS inhomogeneously
doped semiconductor from which we demonstrate that electric field
gradients indeed can play a crucial role in the propagation of spin
polarized electrons. Our findings may in particular highlight a
possible difficulty in obtaining efficient spin polarized
propagation at semiconductor interfaces between low-to-high doping
concentration regions.

In the following we will show a derivation of the drift-diffusion
model leading to eq. (2) (Section 2), followed by our numerical
model (Section 3), and a discussion of our results (Section 4).

\section{Drift-diffusion description of spin transport}

The current for spin-up and spin-down electrons can be written as
\begin{equation}
{\mathbf j}_{\uparrow(\downarrow)} =
\sigma_{\uparrow(\downarrow)}{\mathbf E}+ eD_{\uparrow (\downarrow)}
\nabla n_{\uparrow(\downarrow)}~, \label{DD}
\end{equation} where
$D_{\uparrow(\downarrow)}$ and $\sigma_{\uparrow(\downarrow)}$ are
the diffusion constants and conductivities for the spin-up(down)
species. Here, the conductivities refer to
$\sigma_{\uparrow(\downarrow)}=en_{\uparrow(\downarrow)}\mu_{\uparrow(\downarrow)}$, in terms of the mobilities $\mu_{\uparrow(\downarrow)}$ and densities $n_{\uparrow(\downarrow)}$ of spin-up(down) electrons. The continuity equations for the two species require that
\begin{equation}
\frac{\partial n_{\uparrow(\downarrow)}}{\partial t} = -
\frac{n_{\uparrow(\downarrow)}}{\tau_{\uparrow\downarrow(\downarrow\uparrow)}}
+
\frac{n_{\downarrow(\uparrow)}}{\tau_{\downarrow\uparrow(\uparrow\downarrow)}}
+ \frac{1}{e}\nabla \cdot {\mathbf j}_{\uparrow(\downarrow)}~,
\label{continuity}
\end{equation} where
$\tau_{\uparrow\downarrow}(\tau_{\downarrow\uparrow})$ are the
spin-flip scattering rates. In steady-state eqs.
(\ref{DD},\ref{continuity}) result in
\begin{equation}
\nabla \sigma_{\uparrow(\downarrow)}\cdot {\mathbf E} + \sigma_{\uparrow(\downarrow)}\nabla \cdot {\mathbf E} +eD_{\uparrow(\downarrow)}\nabla^{2} n_{\uparrow(\downarrow)} =  -\frac{en_{\uparrow(\downarrow)}}{\tau_{\uparrow\downarrow(\downarrow\uparrow)}} + \frac{en_{\downarrow(\uparrow)}}{\tau_{\downarrow\uparrow(\uparrow\downarrow)}}~. \label{DDc}
\end{equation} For a NMS
$\mu_{\uparrow}=\mu_{\downarrow}$, and
$D_{\uparrow}=D_{\downarrow}$. Multiplying eq.
(\ref{DDc}) with $\sigma_{\downarrow}(\sigma_{\downarrow})$ 
and substracting them from each other we arrive
to the following expression for the spin density imbalance
$\delta_{\uparrow\downarrow}=n_{\uparrow}-n_{\downarrow}$
\begin{equation}
\nabla^{2} \delta_{\uparrow\downarrow} + \frac{e{\mathbf
E}}{k_{B}T}\nabla \delta_{\uparrow\downarrow} + \frac{e{\mathbf
E}}{2k_{B}T}\delta_{\uparrow\downarrow} \nabla \cdot {\mathbf E} -
\frac{\delta_{\uparrow\downarrow}}{L_{s}}=0~,
\label{DDimbalance}
\end{equation} where we have used the Einstein
relation $\mu=eD/k_{B}T$, $\tau_{s}$ is the spin relaxation time
$\tau_{s}^{-1}=\tau_{\uparrow\downarrow}^{-1}+\tau_{\downarrow\uparrow}^{-1}$
($\tau_{\uparrow\downarrow}=\tau_{\downarrow\uparrow}$), and
$L_{s}=\sqrt{D\tau_{s}}$ is the intrinsic spin diffusion length.

Equation (\ref{DDimbalance}) is a drift-diffusion equation for the
spin density imbalance $\delta_{\uparrow\downarrow}$ and contains
terms depending on both the electric field, and the electric field
{\em gradient}. We note that in fact eq. (\ref{DDimbalance})
resembles the corresponding equations in Refs. \cite{yu}, but
with an {\em additional term proportional to} $\nabla \cdot {\mathbf
E}$. Similarly to Refs. \cite{yu} we can use the roots of
the characteristic equation for eq. (\ref{DDimbalance}) to define
up- and down-stream spin diffusion lengths from eq.
(\ref{DDimbalance}) leading up to eq. (\ref{LDU}). Notice that eq.
(\ref{LDU}) is only defined "locally", over a region where $\nabla
\cdot {\mathbf E}$ can be considered constant, and where an
average value of the electric field is used. Note also that the
above equations are also valid for ferromagnetic semiconductors
provided the mobility and diffusion constant are replaced with the
ones corresponding to the minority-spin species\cite{yu,
flattePRL}.

A comparison between eq. (\ref{LDU}) and Refs. \cite{yu} results in that eq. (\ref{LDU}) above
contains an additional term $-e\nabla E/k_{B}T$ in the square root. Naturally, this
term plays an important role in inhomogeneously doped
semiconductors. We have performed
self-consistent calculations and compared them with numerical fits using eq.
(\ref{LDU}) to verify its applicability.

\section{Self-consistent numerical calculations}
\begin{figure}[t]
\scalebox{1}{\epsfig{file=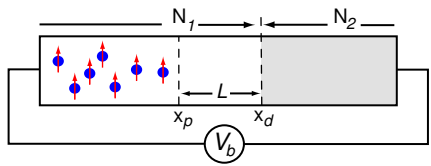}} \caption{
Schematics of the inhomogeneously doped DMS-NMS structure
studied in our calculations. The DMS extends for $x<x_{p}$, whereas
the doping interface occurs at $x=x_{d}$.} \label{fig:1}
\end{figure}
Our self-consistent approach is based on the solution of the
Boltzmann and Poisson equations using numerical methods developed in
Ref.\ \cite{csontosPHYSICAE}. Working in a one-dimensional model (choosing
the $x$ direction), the transport of spin-polarized electrons is
described by two BTE equations according to
\begin{equation}
\label{BTE} -\frac{eE}{m^{\ast}}\frac{\partial f_{\uparrow
(\downarrow)}}{\partial v} + v \frac{\partial
f_{\uparrow(\downarrow)}}{\partial x} =
-\frac{f_{\uparrow(\downarrow)}-f^{0}_{\uparrow(\downarrow)}}{
\tau_{m}} - \frac{f_{\uparrow (\downarrow)}-f_{\downarrow
(\uparrow)}} {\tau_{\uparrow \downarrow (\downarrow \uparrow)}}~,
\end{equation}
where $f_{\uparrow(\downarrow)}$ is the electron distribution for
the spin-up (down) electrons,  $\tau_{m}$ is the momentum relaxation
time, and $1/\tau_{\uparrow \downarrow}$ ($1/\tau_{\downarrow
\uparrow}$) is the scattering rate for spin-up (down)
electrons. The first term on the right-hand side of eq.\ (\ref{BTE})
describes the relaxation of each nonequilibrium spin distribution to
a local equilibrium (spin-dependent), normalized electron
distribution function $f^{0}_{\uparrow (\downarrow)}= n_{\uparrow
(\downarrow)}(\mathbf{r}) \sqrt{\frac{m^{\ast}}{2\pi k_{B}T}}
\exp{(-m\mathbf{v}^{2}/2k_{B}T)}$, where $T=300$ K is the lattice
temperature. The last term in eq.\ (\ref{BTE}) describes the
relaxation of the spin polarization. From the distribution function
$f_{\uparrow (\downarrow)}$ we calculate the local spin density
according to $n_{\uparrow (\downarrow)}(x)= \int f_{\uparrow
(\downarrow)}(r,v)dv$, the total charge density,
$n=n_{\uparrow}+n_{\downarrow}$, and the spin density imbalance,
$\delta_{\uparrow\downarrow}=n_{\uparrow}- n_{\downarrow}$. In order
to take into account inhomogeneous charge distributions and electric
fields we solve eq. (\ref{BTE}) together with the Poisson equation
${\mathbf \nabla} \cdot
{\mathbf E} = e(N_{D} - n_{\uparrow} - n_{\downarrow})/(\varepsilon \varepsilon_{0})$
where $\varepsilon$ is the dielectric constant and $N_{D}$ is the
donor concentration profile, using finite difference and relaxation
methods\cite{csontosPHYSICAE}.

\begin{figure}[b]
\scalebox{0.8}{\epsfig{file=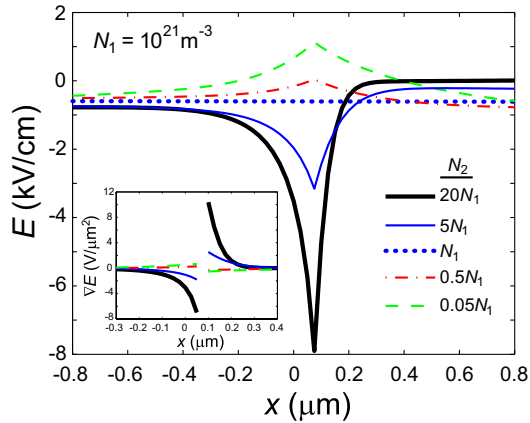}}
\caption{Electric field and electric field gradient (inset) profile
for different doping concentrations $N_{2}=0.05,0.5,1,5,20N_{1}$.}
\label{fig:2}
\end{figure}
\section{Results and discussion}
We have performed self-consistent, numerical calculations for an
inhomogeneously doped DMS-NMS structure schematically depicted in
Fig.\ \ref{fig:1} using $T=300$ K,
$m^{\ast}=0.067m_{0}$, $\tau_{m}=0.1$ ps, $\tau_{s}=0.5$ ns, 
$N_{1}=10^{-21}$ m$^{-3}$, $L=0.2$ $\mu$m and $V_{b}=-0.3$ V.

At the interface between the two doping regions with concentrations
$N_{1}$ and $N_{2}$ a built-in field is expected due to the
diffusion of carriers from the region with higher to the region with
lower doping concentration. This is illustrated in Fig. \ref{fig:2}
where we show the electric field profile, as well as the gradient of
the electric field (inset) around the interfacial regions, for
different doping concentrations $N_{2}=0.05,0.5,1,5,20N_{1}$. The
sign and amplitude of the electric field gradients naturally depend
strongly on the doping concentration $N_{2}$ on a length scale of
the order of the screening length.

While it is not possible to use eq. (\ref{LDU}) directly for a
quantitative analysis of the spin transport properties, one can make
preliminary estimates of the order of the spin relaxation length
around a specific region using linear approximations of the electric
field. For example, using a linear fit of $E$ for $-0.1<x<0.1$
$\mu$m for the $N_{2}=20N_{1}$ curve yields $\nabla E\approx 2.8$
V/$\mu$m$^{2}$. Furthermore, using a mean value of $E\approx 2.9$ kV/cm
and $L_{s}\approx 1.8$ $\mu$m we obtain from eq. (\ref{LDU}) a spin
relaxation length value $L_{D}^{\prime}\approx 0.16$ $\mu$m along
the (charge) direction of transport. Similarly, it is found that
$L_{U}^{\prime}\approx 0.06$ $\mu$m. In comparison, an evaluation of
eq. (\ref{YF}), i.e. without the electric field gradient term, using
the same value of the electric field yields $L_{D}\approx 36$
$\mu$m, and $L_{U}\approx 0.09$ $\mu$m. Thus, the spin relaxation
length in the direction of the charge flow is predicted to be
strongly influenced by the inhomogeneous electric field around the
doping interface.
\begin{figure}[t]
\scalebox{0.8}{\epsfig{file=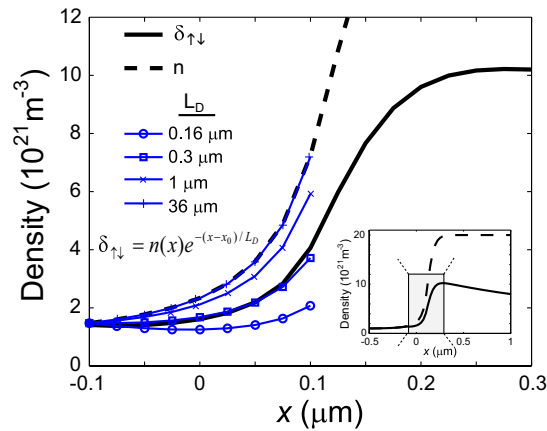}}
\caption{Total charge density (dashed line) and spin density
imbalance (solid line) profiles calculated for $N_{2}=20N_{1}$.
Additional curves for $-0.1<x<0.1$ $\mu$m correspond to
$\delta_{\uparrow\downarrow}=n(x)\exp[-(x+0.1)/L_{D}]$ for different
$L_{D}$ as described in the text. Inset shows $n$ and
$\delta_{\uparrow\downarrow}$ for a larger region.} \label{fig:3}
\end{figure}

In Fig.\ \ref{fig:3}, the numerically calculated spin density
imbalance, $\delta_{\uparrow\downarrow}$, and total charge density,
$n$, are shown for the $N_{2}=20N_{1}$ structure considered in the
above estimates. The total charge density (dashed line) increases
monotonically as expected within a screening length from the low to
the high doping concentration region. The spin density imbalance
(solid line) follows a similar increase, but is, however,
{\em significantly suppressed} such that only a maximum of 0.5$N_{2}$ is
reached around the interface region, beyond which
$\delta_{\uparrow\downarrow}$ decreases exponentially. The observed 
suppression is a consequence of the inhomogeneous electric fields.

A crude estimate of the validity of the values for $L_{D}^{\prime}$
can be obtained by comparing our numerical calculations with
$\delta_{\uparrow\downarrow}=n(x)\exp[-(x-x_{0})/L_{D}^{\prime}]$.
In Fig.\ \ref{fig:3} we show the resulting fits for
$L_{D}^{\prime}=0.16,0.3,1$ and 36 $\mu$m, the first and latter
values corresponding to the previously estimated values for
$L_{D}^{\prime}$ and $L_{D}$ obtained from the numerically
calculated electric field profiles. We see that the higher values of
$L_{D}$ indeed yield that $\delta_{\uparrow\downarrow}\approx n$. In
contrast, the $L_{D}$ values 0.16 and 0.3 $\mu$m, which are of the
order of the screening length, yield better fits to the numerically
calculated data.

Our results and analysis shows that inhomogeneous electric fields
can strongly influence the spin relaxation length, and that
significant spin polarization suppression can occur as a result
thereof. We have also
high-lighted a possible mechanism that may inhibit propagation of
spin polarized electrons from low-to-high regions of doping
concentrations, such as encountered in future semiconductor
spintronic devices.

This work was supported by the Indiana 21$^{{\rm st}}$ Century Research and
Technology Fund. Numerical calculations were performed using the
facilities at the Center for Computational Nanoscience at Ball State
University.

\end{document}